\documentclass[12pt,preprint]{aastex}
\usepackage{bm}

\def\CQG{Class. Quantum Gravity}

\def\FP{Fortschr. Physik}
\def\GRG{Gen. Relativity and Gravitation}

\def\JHEP{JHEP}

\def\MPL{Mod. Phys. Lett.}
 
\def\NC{Il Nuovo Cimento}
\def\NPB{Nucl. Phys. B}
\def\PL{Phys. Lett. }

\newcommand{\be}{\begin{equation}}
\newcommand{\ee}{\end{equation}}
\newcommand{\beqa}{\begin{eqnarray}}
\newcommand{\eeqa}{\end{eqnarray}}
\shorttitle{Future Constrains On Chaplygin Gas Models.}
\slugcomment{Preprint DF/IST-2.2003}

\begin{document}

\title{Expected constraints on the generalized Chaplygin equation of state from
future supernova experiments and gravitational lensing statistics.}

\author{P.T. Silva}\email{paptms@ist.utl.pt} \and \author{O. Bertolami}
\email{orfeu@cosmos.ist.utl.pt}
\affil{ Instituto Superior T\'ecnico, Departamento de F\'\i sica.\\Av. Rovisco
Pais 1, 1049-001 Lisboa, Portugal}

\begin{abstract}

This paper aims to study the use of future SNAP data together with the result
of searches for strong gravitational lenses in future large quasar surveys to
constrain the Generalized  Chaplygin Gas (GCG) model, with equation of state
$p=-A/\rho^\alpha$, where $A$ is a positive constant and $0<\alpha\le1$. The
GCG is considered as a possible unification scheme for dark matter-dark energy.
It is found that both experiments should be able to place important 
constraints on the model, especially when both tests are used together.

\end{abstract}

\keywords{cosmology: cosmological parameters, dark matter --- equation of state
--- gravitational lensing}

\maketitle

\section{Introduction}

In the last few years there has been mounting evidence that not only visible
matter is a small component of the observable Universe, but that matter itself,
dark or visible, makes up only a fraction of the cosmic energy budget. Indeed,
SNe Ia experiments indicate that the Universe is expanding in an 
accelerated fashion \citep{Riess1998,Garnavich1998,Perlmutter1999}, and
together with nucleosynthesis constraints 
\citep{Burles2001}, and the CMBR power spectrum \citep{Balbi2000,
deBernardis2000, Jaffe2001} it is possible to arrive to a concordance where the
Universe is made of about $5\%$ of baryonic matter, $25\%$ of dark matter and
$70\%$ of dark energy, a negative pressure component. These results are also in
agreement with large scale structure \citep{Peacock2001}, and independent
determinations of the matter density \citep{Bahcall1998, Carlberg1998,
Turner2000}. The nature of the dark matter and of the dark energy is not well
established; dark matter candidates include axions,
neutralinos and possibly a self-interacting scalar-field (see e.g. Bento, 
Bertolami, \& Rosenfeld 2001 and references therein), while the most obvious
candidate for the dark energy is the vacuum energy, or an uncanceled
cosmological constant \citep{Bento1999, Bento2001b}. Another possibility is a 
dynamical vacuum  \citep{Bronstein1933, Bertolami1986a, Bertolami1986b,
Ozer1987} or quintessence. Quintessence models most often involve a single
scalar field 
\citep{Ratra1988a, Ratra1988b, Wetterich1988, Caldwell1998, Ferreira1998,
Zlatev1999, Binetruy1999, Kim1999, Uzan1999,Amendola1999, Albrecht2000,
Bertolami2000,Banerjee2001a,Banerjee2001b,SenSen2001,Sen2001} or two coupled
fields  \citep{Fujii2000, Masiero2000, Bento2002c}. However, these models are
afflicted with a fine tuning problem to explain the cosmic coincidence problem,
that is, why did the dark  energy start to dominate the cosmological evolution
only fairly recently.

Recently a new possibility has been suggested in \citet{Kamenshchik2001},
Bili\'c, Tupper, \& Viollier (2002), and Bento, Bertolami, \& Sen (2002a),
which uses an exotic equation of state, and considers the evolution 
of the equation of state of the background fluid instead of a quintessence
potential. This is
achieved by using a background fluid, the generalized Chaplygin gas (GCG),
described by the equation of state \citep{Kamenshchik2001,Bento2002a}
\be
p_{ch}=-{A\over\rho_{ch}^\alpha}\;.
\label{EqState}
\ee
The case $\alpha=1$ is called the Chaplygin equation of state, $p_{ch}=-A/
\rho_{ch}$, while when $\alpha=0$, the model behaves as the flat $\Lambda CDM$,
as discussed  in Section II. The quantity $A$ is related to the present
day Chaplygin adiabatic sound speed through $v_s^2=\alpha A/\rho_{ch,0}$, where
$\rho_{ch,0}$ is the GCG density at present. This model allows for an
unification of dark energy and dark matter and admits a brane interpretation
\citep{Bento2002a}. Furthermore, analyses indicate
that it may be accommodated within the standard structure formation scenario
\citep{Bilic2002,Bento2002a}. These results have been challenged in
\citet{Sandvik2002}, however their analysis does not take into account
the effect of baryons, which is expected and shown to be important
\citep{Beca2003}, and makes the GCG model compatible with the 2DF mass
power spectrum. Also, \citet{Sandvik2002} analysis is based on the
linear treatment of pertubations close to the present time, thus neglecting
any non-linear effects. Thus we consider that a more careful and thorough 
analysis must be done before ruling out the GCG model on the grounds of 
its effects in the matter power spectrum or structure formation.

Since the GCG exhibited great potentialities, it has been recently the subject
of great interest, and various attempts have been made to constrain the model
using the available observational data. In \citet{Avelino}
the GCG was tested against the SNe Ia data and the matter power spectrum, and
it is concluded that the GCG model agrees with the data, and interesting
constraints on $A_s=A/\rho_{ch,0}^{1+\alpha}$ are found for fixed values of
$\alpha$. Also, the value $A_s=1$, corresponding to a pure cosmological
constant, is strongly ruled out by data. The limits were somewhat degenerate
mainly because three parameters, $\alpha$, $A_s$,
and the density relative to critical  of the Chaplygin gas, $\Omega_{ch}$,
were considered. In \citet{Dev1} tests on the age of the Universe and strong
lensing statistics were used to constrain the Chaplygin equation  of state
($\alpha=1$), and it is found that
the available data agrees with a Chaplygin gas model; from the age test 
it is concluded that $A_s\geq0.96$, and from the statistics of lensed
quasars it is concluded that $A_s\geq0.72$. The age test together with SNe
distance measurements were used to constrain the generalized 
Chaplygin equation of state in a flat Universe made up of the GCG
and baryons, with fixed densities relative to critical,
yielding confidence regions in the $(A_s;\alpha)$ plane (Makler, Oliveira \&
Waga 2002). The
data indicated that $0.6\lesssim A_s \lesssim 0.85$ at  95\% confidence level
(CL) and almost no constraints of interest to parameter $\alpha$. The strongest
constraints were obtained by \citet{Bento2002b} using the CMBR power spectrum
measurements from BOOMERANG \citep{Boomerang} and Archeops \citep{Archeops},
together with  the SNe Ia constraints from \citet{Makler}. It is found that
$0.74\lesssim A_s \lesssim 0.85$, and $\alpha\lesssim0.6$, ruling out the pure
Chaplygin gas model. Using the bound arising from the age of the APM 08279+5255
source, which is $A_s\gtrsim0.81$ (Alcaniz, Jain, \& Dev 2002), fairly tight
constraints are found, namely  $0.81\lesssim A_s\lesssim0.85$, and
$0.2\lesssim\alpha\lesssim0.6$, which also rules out the $\Lambda CDM$ model.
These results were found to agree with the WMAP data \citep{Bento2003}.

The prospect of launching the SNAP satellite\footnote{snap.lbl.gov} will give
rise to a new era of precision cosmology, and together with data from CMBR
experiments and from large quasar and galaxy surveys,
cosmologists will be able to test several models with unprecedented accuracy.
Of course, this increase in accuracy may find its limits  in new
degeneracies, given the growth of free parameters, which in turn will require
the use of more cosmological tests to distinguish competing models. In this
paper we aim to study how well the GCG model can be constrained with future
data from the SNAP satellite, and from strong lensing statistics. We will be
especially interested in the use of the Sloan Digital Sky
Survey\footnote{www.sdss.org} (SDSS) data to constrain cosmological parameters,
and also smaller surveys such as the 2dF\footnote{www.2dfquasar.org}. Even
though it has already been verified that the SNAP satellite will be able to
constrain the model quite accurately \citep{Avelino,Makler}, the tighter
constraints provided by the CMBR \citep{Bento2002b,Bento2003} invite us to
go a step forward as SNe confidence regions depend strongly on the fiducial
model considered. Moreover, we add that the use of gravitational
lensing statistics from future large surveys has not been explored in 
the context of the Chaplygin gas. Thus, the main aim of the our study is
to use all three tests to constrain as much as possible the parameters
of the GCG.

The outline of this paper is as follows. In Section \ref{section2} we explain
how the Chaplygin model closely mimics the $\Lambda CDM$ model, and show that
the case $\alpha=0$ corresponds to the $\Lambda CDM$ model.
In Section \ref{section3} we consider the magnitude versus redshift
cosmological test. In \ref{section3A} we briefly describe the nature of this
test, and in \ref{section3B} we estimate the errors considered. In
\ref{section3C} we describe the method used to obtain the confidence regions
shown in \ref{section3D}. 

Section \ref{section4} is dedicated to strong lensing statistics. In
\ref{section4A} we briefly describe the strong lensing test, in \ref{section4C}
we explain how we use the Fisher matrix formalism to build the confidence
regions that are exhibited in \ref{LensingResults}.

\section{\label{section2}
Generalized Chaplygin gas compared to flat $\Lambda CDM$.}

The cosmological tests considered here depend on the cosmological model through
the comoving distance $r(z,H_0,\bm\Theta)$, where $\bm\Theta$ is a vector
containing the cosmological parameters we are interested in determining, and
$H_0$ being the Hubble constant. For a Friedman-Robertson-Walker (FRW) 
Universe the comoving distance as function of redshift is given by
\be
\label{ComovingD}
r(z,H_0,\Theta)=
\left\{
\begin{array}{ll}
\vspace{.3cm}
\displaystyle \frac{c}{H_0\sqrt{-\Omega_k}}\sin\left(
\int_0^z\frac{H_0\sqrt{-\Omega_k}dz'}
{H(z',H_0,\Theta)}\right)& \Omega_k<0 \\
\vspace{.3cm}
\displaystyle
c\int_0^z\frac{dz'}{H(z',H_0,\Theta)}&\Omega_k=0\\
\displaystyle
\frac{c}{H_0\sqrt{\Omega_k}}\sinh\left(\int_0^z\frac{H_0\sqrt{\Omega_k}
dz'}{H(z',H_0,\Theta)}\right)& \Omega_k>0
\end{array}\right.
\ee
where $\Omega_k=1-\Omega$, $\Omega$ being the total density of the Universe
relative to the critical one. Only flat cosmological models, $\Omega_k=0$, are
considered in this paper.

The expansion rate may be written in terms of the total mass-energy density
as a function of redshift. Recalling the Friedman equation,
\be
H^2(z)=\frac{8\pi G}{3}\rho(z)\;,
\ee
where $\rho(z)$ is the total mass-energy density at redshift $z$, the expansion
rate is given by
\be
H(z)=H_0\left(\frac{\rho(z)}{\rho_0}\right)^{1/2},
\ee
where $\rho_0=\rho(0)$ is the energy density at present.
For  a flat Universe, $\rho_0=\rho_c$, where $\rho_c$ is
the critical density.

For the flat $\Lambda CDM$, ignoring the small radiation energy density
contribution, one has
\be
\label{LCDMDensity}
\rho(z)=\rho_0\left[\Omega_\Lambda+\left(\Omega_{CDM}+\Omega_b\right)(1+z)^3
\right]\;,
\ee
where $\Omega_b$, $\Omega_{CDM}$, $\Omega_\Lambda$ are the baryonic, cold dark
matter and vacuum energy densities relative to critical, respectively. Thus,
the Hubble parameter becomes
\be
\label{LCDMHubble}
H(z)=H_0\left[\Omega_\Lambda+(\Omega_{CDM}+\Omega_b)(1+z)^3\right]^{1/2}\;.
\ee
The GCG unifies dark matter and energy, but does not 
account for the baryons. Therefore in a consistent model of the Universe 
baryons and radiation must be included. The contribution of radiation shall be
ignored, since its effect is not important for the redshifts considered here.
Hence, for a Universe made up of the GCG plus baryons one has
\citep{Bento2002a}
\be
\rho_{ch}(z)=\rho_{b,0}(1+z)^3+\left(A+B(1+z)^{3(1+\alpha)}\right)^{1/
(1+\alpha)}\;.
\ee
where $\rho_{b,0}$ is the baryon density at present, and $B$ is an integration
constant. Using the freedom to define the integration constant $B$, this
expression may be rewritten as
\be
\label{chapDensity}
\rho_{ch}(z)=\rho_{b,0}(1+z)^3+\rho_{ch,0}\left[A_s+(1-A_s)
(1+z)^{3(1+\alpha)}\right]^{1/(1+\alpha)}\;,
\ee
where $\rho_{ch,0}=(A+B)^{1/(1+\alpha)}$ is the present day density of the
Chaplygin gas, and $A_s\equiv A/\rho_{ch,0}^{1+\alpha}$. Thus, for a flat
Universe, the Friedman equation becomes
\be
\label{chapHubble}
H_{ch}(z)=H_0\left\{\Omega_b(1+z)^3
+\Omega_{ch}\left[A_s+(1-A_s)(1+z)^{3(1+\alpha)}
\right]^{1/(1+\alpha)}\right\}^{1/2}\;,
\ee
where $\Omega_{ch}=\rho_{ch,0}/\rho_c$ is the GCG density relative to the 
critical one.

Even though the generalized Chaplygin equation of state looks rather exotic,
the resulting cosmological model behaves in a manner that mimics the $\Lambda 
CDM$ model
\citep{Bento2002a}. First let us consider the very early Universe, when its
size was very small compared to the present one, $a(z)\ll a_0\equiv1$, or
$z\gg0$. The $\Lambda CDM$ behaves as a flat CDM dominated Universe: 
\be
\rho(z\gg0)=\rho_0\left(\Omega_{CDM}+\Omega_b\right)(1+z)^3\;.
\ee
The GCG also behaves as a flat CDM dominated Universe
with a dark matter density $\Omega_{CDM}=\Omega_{ch}(1-A_s)^{1/(1+\alpha)}$,
as may be seen in
\be
\rho_{ch}(z\gg0)=\rho_0\left(\Omega_b+\Omega_{ch}(1-A_s)^{1/(1+\alpha)}
\right)(1+z)^3\;.
\ee
This property ensures that the GCG model is consistent, at large
scales,  with the CDM structure formation scenario.

Consider now the late Universe, when $a\gg a_0$. The $\Lambda CDM$ behaves as a
vacuum dominated Universe,
\be
\rho(a\gg a_0)=\rho_0\Omega_\Lambda\;,
\ee
while the GCG model also behaves as a vacuum dominated Universe, that 
at present has a vacuum density $\Omega_\Lambda=\Omega_{ch }
A_s^{1/(1+\alpha)}$,
\be
\rho_{ch}(a\gg a_0)=\rho_0\Omega_{ch}A_s^{1/(1+\alpha)}\;.
\ee

Between these two limits, the GCG density may be expanded in 
subleading terms, yielding
\be
\rho_{ch}=\rho_0\left\{\Omega_b(1+z)^3
+\Omega_{ch}\left[A_s^{1/(1+\alpha)}
+\frac{1}{1+\alpha}\frac{1-A_s}{A_s^{\alpha/(1+\alpha)}}(1+z)^{3
(1+\alpha)}\right]\right\}\;.
\ee
Thus, between the dust and the De Sitter phases, the GCG behaves like
a two component fluid made up of vacuum energy with a present day density
given by $\Omega_\Lambda=\Omega_{ch}A_s^{1/(1+\alpha)}$, and soft matter
with the equation of state $\rho=\alpha p$, and at present
\be
\Omega_{sm}=\Omega_{ch}\left(\frac{1}{1+\alpha}\right)\frac{1-A_s}{A_s^{\alpha
/(1+\alpha)}}\;.
\ee

Also, making the associations
\be
\begin{array}{ccc}
\Omega_{ch}^{1+\alpha}A_s&\rightarrow&\Omega_\Lambda^{1+\alpha}\;,\\
\\
\Omega_{ch}^{1+\alpha}(1-A_s)&\rightarrow&\Omega_{CDM}^{1+\alpha}\;,
\end{array}
\ee
and using them in Eqs. (\ref{chapDensity}-\ref{chapHubble}), one obtains
\beqa
\rho_{ch}(z)&=&\rho_{b,0}(1+z)^3+\rho_0\left[\Omega_\Lambda^{1+\alpha}+
\Omega_{CDM}^{1+\alpha}(1+z)^{3(1+\alpha)}\right]^{1/(1+\alpha)}\;,
\\
\nonumber\\
H_{ch}(z)&=&H_0\Bigg\{\Omega_b(1+z)^3+\left[\Omega_\Lambda^{1+\alpha}+
\Omega_{CDM}^{1+\alpha}(1+z)^{3(1+\alpha)}\right]^{1/(1+\alpha)}
\Bigg\}^{1/2}\;,
\\
\nonumber\\
\label{chapFlat}
&&\left[\Omega_\Lambda^{1+\alpha}+\Omega_{CDM}^{1+\alpha}\right]^{1/(1+\alpha)}
+\Omega_b=1\;,
\eeqa
which allows one to see that $\alpha=0$ is identical to the flat
$\Lambda CDM$ model.

\section{\label{section3}Magnitude versus redshift cosmological test.}

\subsection{\label{section3A}Description of the test.}

The magnitude versus redshift test explores the dependence of the distance
modulus of a source on its redshift. For a source with redshift $z$, its 
apparent magnitude is related to its absolute magnitude through
\be
m(z)=M+5\log d_L(z,H_0,\bm{\Theta}) + 25\;,
\ee
where $\bm\Theta$ represents the several cosmological parameters we are 
interested in, $d_L$ is the luminosity distance, $d_L=(1+z)r$, measured 
in units of 10 pc, and $r$ is the comoving distance, Eq. (\ref{ComovingD}). 
The luminosity distance is then given by
\be
\label{LumDis}
d_L(z,H_0,\Theta)=
\left\{
\begin{array}{ll}
\vspace{.3cm}
\displaystyle\frac{(1+z)c}{H_0\sqrt{-\Omega_k}}\sin\left(
\int_0^z\frac{H_0\sqrt{-\Omega_k}dz'}
{H(z',H_0,\Theta)}\right) & \Omega_k<0\;, \\
\vspace{.3cm}
\displaystyle
(1+z)c\int_0^z\frac{dz'}{H(z',H_0,\Theta)} & \Omega_k=0\;,\\
\displaystyle
\frac{(1+z)c}{H_0\sqrt{\Omega_k}}\sinh\left(\int_0^z\frac{H_0\sqrt{\Omega_k}
dz'}{H(z',H_0,\Theta)}\right) & \Omega_k>0\;.
\end{array}\right.
\ee

Defining a new quantity $D_L=H_0d_L$ one may group the terms that depend on
the absolute magnitude and $H_0$. The apparent magnitude then reads
\beqa
m(z,H_0,\Theta)&=&M-5\log(H_0)+25-5\log D_L(z,\Theta)\nonumber\\
\label{mz}
               &=&\mathcal{M}-5\log D_L(z,\Theta)\;.
\eeqa
The quantity $\mathcal{M}=M-5\log H_0 +25$ is usually called zero point 
value or {\it intercept} since it it the value of the apparent magnitude at the
point where $\log D_L=0$. This quantity can be measured using low redshift SNe,
or be treated  as a  statistical nuisance that is marginalized
\citep{Perlmutter}. We shall follow \citet{Goliath} and assume either an exact
knowledge of $\mathcal{M}$ or no prior knowledge of it. This allows us to
concentrate on the study of the parameters that are relevant to us.

\subsection{\label{section3B}Error estimates for one year of SNAP data.}

In the next subsections we aim to study the ability of the SNAP mission to test
the GCG equation of state. We shall use the expected error estimates for the
SNAP satellite from \citet{Albrecht}, and consider that the systematic 
errors for the aparent magnitude $m$ are given by
\be
\sigma_{sys}(z)=\frac{0.02}{1.5}z\;,
\ee
which are measured in magnitudes such that at $z=1.5$ the systematic error is 
$0.02$ magnitudes,  while the statistical errors for $m$ are estimated to
be $\sigma_{sta}=0.15$ magnitudes. We place the SNe in bins of
width $\Delta z\approx0.05$. We then sum both kind of errors quadratically
\be
\sigma_{mag} (z_i)= \sqrt{\sigma^2_{sys}(z_i)+\frac{\sigma^2_{sta}}{n_i}}\quad,
\ee
where $n_i$ is the number of supernovae in the redshift bin. The distribution
of supernovae in each redshift bin is the same considered in \citet{Albrecht},
and is shown in Table \ref{SNAPerrors}.


\subsection{\label{section3C}Confidence regions for SNe Ia tests.}

To build the confidence regions a fiducial model corresponding to a vector
$\bm\Theta_{true}$ is chosen, and log-likelihood functions $\chi^2$ are 
calculated based on hypothetical magnitude measurements at the various
redshifts. This function is given by
\be
\chi^2(\mathcal{M},\bm\Theta)=\sum_{z_i=0}^{z_{max}}\frac{\left[m(z,
\mathcal{M},\bm\Theta)-m(z,\mathcal{M},\bm\Theta_{theory})\right]^2}{
\sigma^2(z)}\;,
\ee
where the sum is made over all redshift bins and $m(z,\mathcal{M},\bm\Theta)$
is defined as in Eq. (\ref{mz}).

The $\chi^2$ function is directly related to the maximum likelihood
estimator of a normal random variable, $L=\exp(-\chi^2/2)$. Suppose one
wants to impose a Gaussian prior to one of the parameters being measured,
$\theta$, centered around $\theta_0$, with variance $\sigma^2_\theta$. Using
the Bayes theorem one finds that
\be
L=\exp(-\frac{1}{2}\chi^2)\exp\left(\frac{(\theta-\theta_0)^2}
{2\sigma^2_{\theta}}\right)\;.
 \ee
Thus, the $\chi^2$ function is changed to
\be
 \chi^2 \rightarrow \chi^2+\frac{(\theta-\theta_0)^2}{\sigma^2_{\theta}} 
\;.
\ee

If one aims to impose a prior on a given parameter one is not measuring,
one has to integrate the parameter out of the likelihood function using
its probability density function, $p(\theta)$, as prior, that is
\be
\chi^2_{\textrm{prior on }\theta}=-2\ln\int_{-\infty}^{+\infty}d\theta\exp
\left(-\frac{1}{2}\chi^2\right)p(\theta)\;.
\ee

A special case is the marginalization over the zero point $\mathcal{M}$. When
this zero point is exactly known it cancels out from the $\chi^2$ 
expression, and the modified log-likelihood function $\hat{\chi}^2$
\citep{Goliath} is obtained,
\be
\widehat{\chi}^2=\sum_{i=1}^n\frac{\Delta^2}{\sigma_i^2}\;,
\ee
where
\be
\Delta=5\log_{10}[D_L(z,\bm\Theta)]-5\log_{10}[D_L(z,\bm\Theta_{true})]\;.
\ee

Assuming no prior at all about $\mathcal{M}$, one has to integrate it out
of the $\chi^2$ function, obtaining a modified log-likelihood function
$\widetilde{\chi}^{\,2}$:
\begin{eqnarray}
\widetilde{\chi}^{\,2} & = & -2\ln\int_{-\infty}^{+\infty}d\mathcal{M}\exp
\left(-\frac{1}{2}\chi^2\right)\nonumber\\
 & = & \widehat{\chi}^2-\frac{B^2}{C}+\ln\left(\frac{C}{2\pi}\right)\;,
\end{eqnarray}
where
\be
B=\sum_{i=1}^n\frac{\Delta}{\sigma_i^2}\;,
\ee
\be
C=\sum_{i=1}^n\frac{1}{\sigma_i^2}\;.
\ee

\subsection{\label{section3D}Expected Confidence regions for 1 year of SNAP
data.}

The values used for each fiducial model are shown in Table
\ref{FiducialModels}. As mentioned above, the use of SNAP data to constrain
the GCG model was already considered in \citet{Makler} and \citet{Avelino},
but our analysis differs these ones in three points. Firstly, the \citet{Avelino} 
analysis is done in the $(\Omega_{ch},A_s)$ plane for fixed values of
$\alpha$, while we consider $\Omega_{ch}$ fixed, and find constraints
in the $(A_s,\alpha)$ plane. Secondly, in \citet{Makler} only a $\Lambda
CDM$ fiducial model is considered, essentially with the purpose of 
testing whether SNAP will be able to rule out the GCG model, noting 
that it will be possible to distinguish between a $\Lambda CDM$ and the
Chaplygin gas ($\alpha=1$) model. Our analysis not only considers the capability
of ruling out the GCG model, via our study of Model I, but also considers 
the precision of the constrains one may find for both our fiducial models.
The inclusion of this second
fiducial model is important, as it corresponds to the center point of the 
range allowed by CMBR \citep{Bento2002b} and the age test \citep{Dev2}.
Thirdly, we are not only interested in the use of SNAP to constrain the
model, but also the use of strong gravitational lensing statistics, and
especially to study what constraints might arise when both probes are
used together. Throughout the paper we will also use the CMBR constraints 
from \citet{Bento2002b}, essentially with two purposes. Firstly to 
compare these available constraints with the future ones, to see whether
future tests will be able to improve the available constrains or merely 
confirm them, and secondly as a third test that might be used to improve
the confidence regions.

\begin{figure}
\centering
\plotone{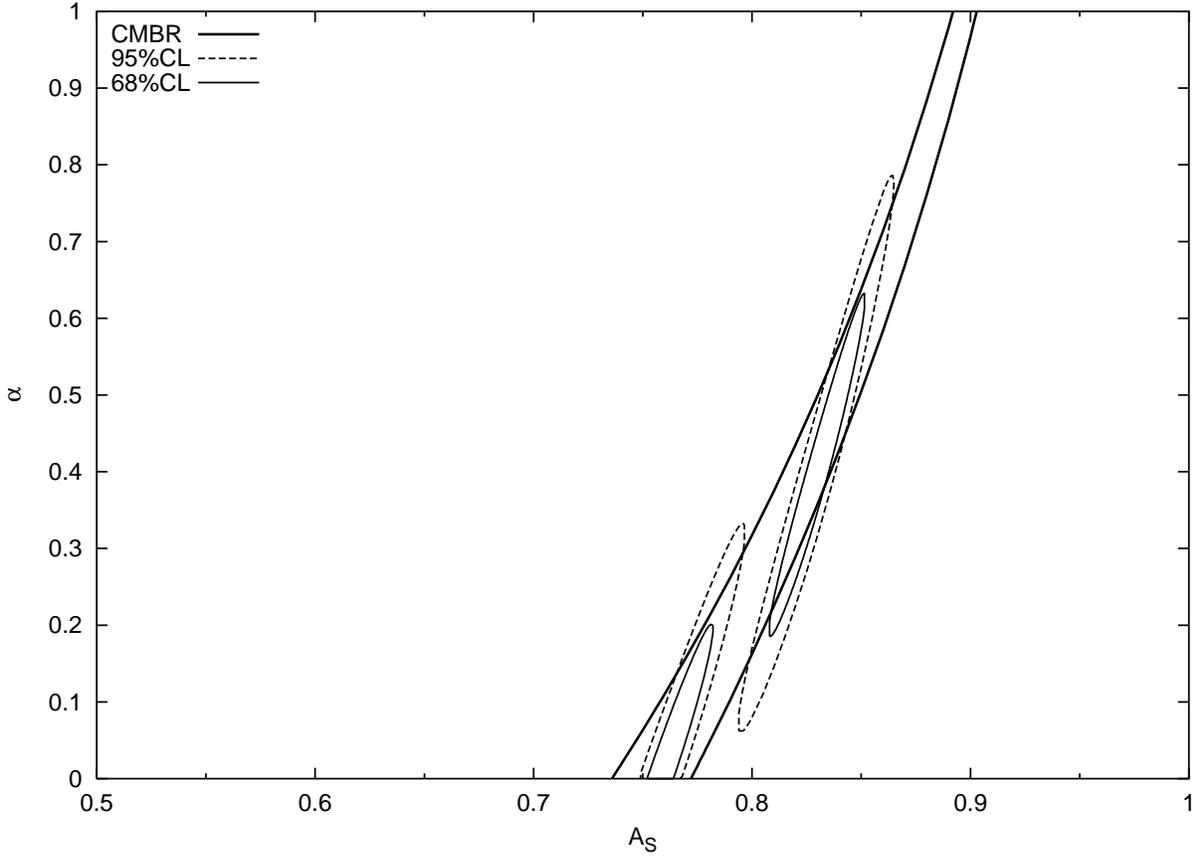}\\
\caption{Expected SNAP confidence regions assuming precise knowledge of 
$\mathcal{M}$ plus the CMBR constraints from \citet{Bento2002b}.}
\label{Lum+CMB_ExactM}
\end{figure}

\begin{figure}
\centering
\plotone{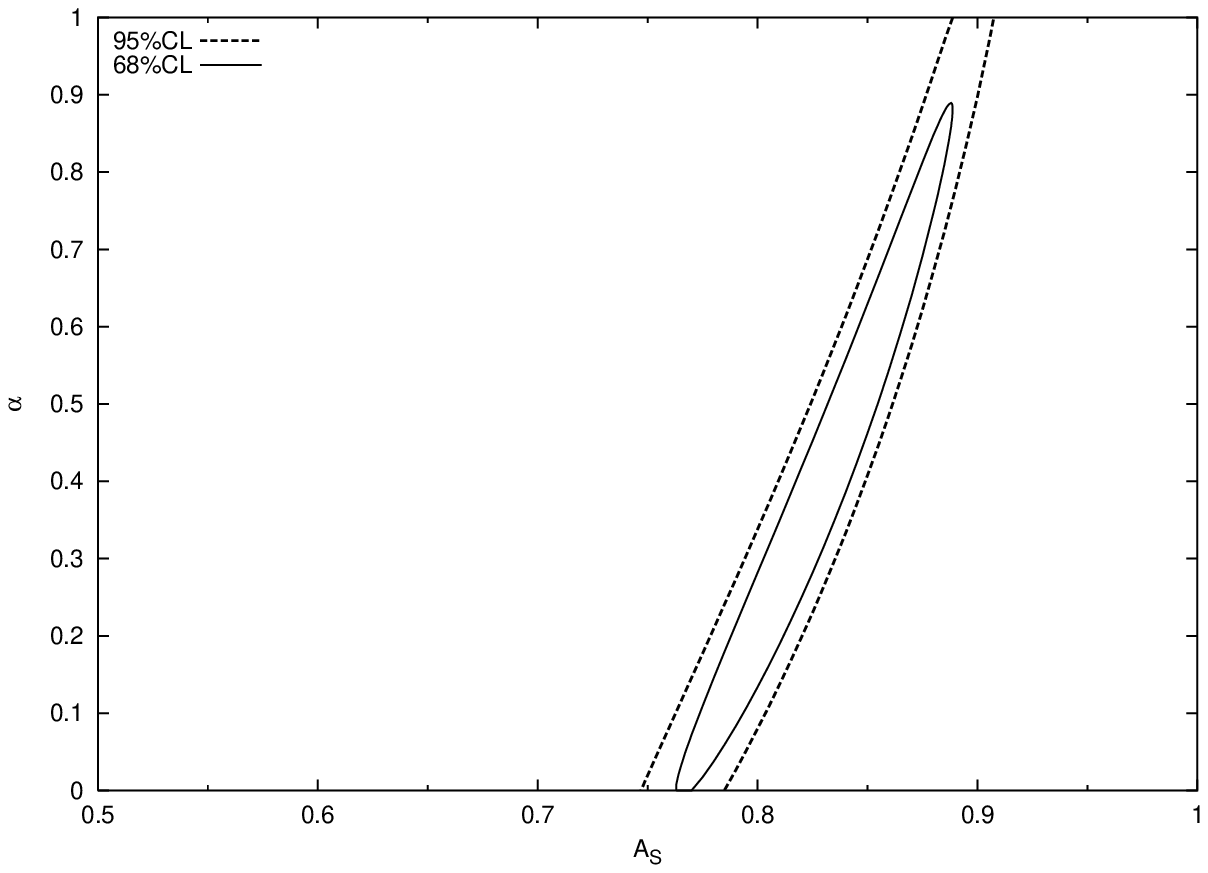}\\
\caption{Expected SNAP confidence regions for Model I assuming no knowledge of 
$\mathcal{M}$.}
\label{Model1_LumNoM}
\end{figure}

\begin{figure}
\centering
\plotone{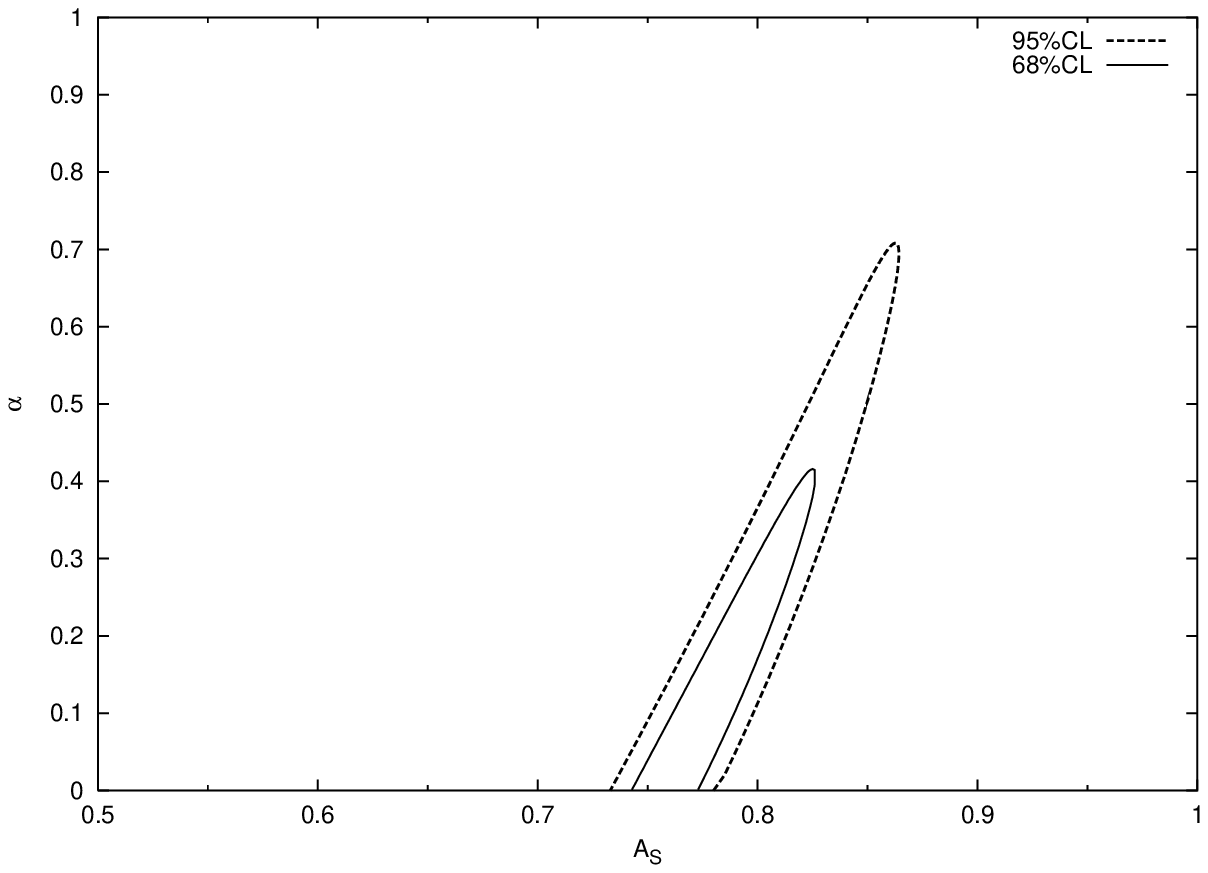}\\
\caption{Expected SNAP confidence regions for Model II assuming no knowledge of
$\mathcal{M}$.}
\label{Model2_LumNoM}
\end{figure}

The expected confidence regions that are obtained are shown in Figures 
\ref{Lum+CMB_ExactM}-\ref{Model2_LumNoM}. In all cases a fixed baryon density
with value $\Omega_b=0.05$ is used. Two fiducial models were considered. Model
I corresponds to the center value of the parameter range found by 
\citet{Bento2002b}, $(A_s,\alpha)=(0.83,0.4)$. Model II corresponds to a
flat $\Lambda CDM$ model with $\Omega_\Lambda=0.72$,
$\Omega_{CDM}=0.23$, and $\Omega_b=0.05$. Using the associations given in
section 3.1, this corresponds to $(A_s,\alpha)=(0.758,0)$.
Both sets of values are compatible with the CMBR data 
\citep{Bento2002b} and available SNIa data \citep{Makler}. The set
corresponding to the $\Lambda CDM$ fiducial model is ruled out by the age
estimate of the APM 08279-5255 source \citep{Dev2}, but it was included to
illustrate the effect of choosing a fiducial model. 


The first conclusion one draws is that in order to achieve some precision, a 
good estimate of the intercept $\mathcal{M}$ is required. Without imposing
prior knowledge of $\mathcal{M}$  it is possible to distinguish only
between a $\Lambda CDM$ model ($\alpha=0$) and a generalized Chaplygin model
with $\alpha\gtrsim0.75$ at 95\% confidence level (CL), or with
$\alpha\gtrsim0.4$ at 68\% CL, as shown in  Figures \ref{Model1_LumNoM}
and \ref{Model2_LumNoM}. For Model I (Figure \ref{Model1_LumNoM}) the
situation is even
worse, and without a good knowledge of the intercept it is only possible to
distinguish between models with $\alpha\gtrsim0.9$ at 68\% CL.
In either case the pure Chaplygin gas ($\alpha=1$) is ruled  out with 68\% CL.
To distinguish between both fiducial models a good estimate of $\mathcal{M}$ is
necessary. 

In the previous paragraph the ability of the test to distinguish
between models was considered. Next we consider whether it is possible to
improve on the limits imposed by the CMBR \citep{Bento2002b}, available SNe
\citep{Makler} and the age test \citep{Dev2}. We find that the answer depends
on the fiducial model considered. For Model II it is possible to confirm or to
reject the present limits. Indeed, using CMBR constraints from
\citet{Bento2002b}, Figure \ref{Lum+CMB_ExactM}, leads to a smaller parameter
region, but it allows no  more than to assert that  
$0.2\lesssim\alpha\lesssim0.65$ and $0.81\lesssim A_S\lesssim 0.85$ at 68\% CL,
 or $0.15\lesssim\alpha\lesssim0.75$ and $0.79\lesssim A_S\lesssim 0.87$ at 
95\% CL.

For Model I, the situation is different. This model is not compatible with 
the age estimate of the APM 08279-5255 source \citep{Dev2}, therefore
the comparison must be made with the allowed parameter range from available
SNe data and the CMBR constraints from \citet{Bento2002b}. From only these two
tests one can establish that $0.74\lesssim A_s\lesssim0.85$, and
 $0\lesssim\alpha\lesssim0.6$ \citep{Bento2002b,Makler}. For this fiducial
model, the SNAP data will  allow for tighter constraints, yielding
$0.75\lesssim A_S\lesssim0.80$ and $\alpha\lesssim0.35$ at 95\% CL. The
use of additional CMBR constraints from \citet{Bento2002b} will allow
further constraining $\alpha$ to $\alpha\lesssim0.3$ at 95\% CL.



\section{\label{section4}Strong lensing probability.}

\subsection{\label{section4A}Description of the test.}

We start the description  of the strong lensing probability test by setting
the assumptions which are most commonly used in gravitational lensing
statistics, namely:
\begin{itemize}
\item{Lensing galaxies are described by singular isothermal sphere (SIS)
profiles.}
\item{There is no evolution in the population of lensing galaxies.}
\item{The Tully-Fisher and Faber-Jackson relations are independent of 
redshift.}
\end{itemize}
Under these assumptions, in a flat FRW Universe the probability that a 
source at redshift $z_S$ will be lensed by a galaxy is given by 
(Turner, Ostriker, \& Gott 1984; Gott, Park, \& Lee 1989)
\be
\label{F}
\tau=F\frac{r^3(z_S)}{30R_0^3}\;,
\ee
where $R_0=c/H_0$, $r(z)$ is the comoving distance, Eq. (\ref{ComovingD}), and
$F$ is the dimensionless parameter
\be
F=\frac{16\pi^3}{cH_0^3}<n_0\sigma_\|^4>\;,
\ee
$n_0$ being the number density of galaxies at present and $\sigma_\|$ is
their velocity dispersion. The average is computed over all lensing galaxies
using the Tully-Fisher and Faber-Jackson relations together with the galaxy
luminosity function \citep{FT}. We shall use the value estimated in 
\citet{Cooray1}, $F=0.026$. The information concerning the  cosmological model
enters through the comoving distance.

\subsection{\label{section4B}Magnification bias and selection effects.}

To sucessfully use strong lensing statistics as a tool to test cosmological 
models, a detailed account of possible systematic effects is required. Here we
provide a brief discussion of some effects and how to quantify them.

One of the most important systematic effect is the magnification bias
\citep{TOG}. It affects magnitude limited surveys,
since gravitationally lensed sources are preferentially
included in the survey. This bias may be quite large, depending on the quasar
luminosity function. Other important biases are the selection effects due to
the limitations on the dynamical range (magnitude differences between the
lensed images), resolution limitations and the presence of confusing sources
such as stars   \citep{Kochanek91,Kochanek93a}. Besides these there is the
issue of selection effects associated with quasar surveys (which tends to
misrepresent the quasar population) such as the luminosity of the lensing
galaxy (which tends to overwhelm the quasar luminosity and exclude it from the
catalogue) and the reddening due to the lensing galaxy (which may reduce the
average magnification bias, increase flux ratios and absorb single images)
\citep{Kochanek91}.

Effects due to the lensing galaxy luminosity are expected to be marginal for
bright quasar surveys, but for deep (low luminosity cut-off) surveys such 
as the SDSS the effect might be statistically dominant. Following
\citet{CoorayHut}, we shall not include this correction. A correct estimate
of these effects requires an accurate characterization of the apparent
magnitudes of lensing galaxies as function of redshift \citep{Kochanek91}. 
Since there still is no consensus on how to quantify
the reddening effect \citep{malhotra,Falco}, we shall follow, as before,
\citet{CoorayHut} and disregard this effect too.

The effect of confusing sources should be negligible if the
lensing survey magnitude range is not excessively large ($\Delta m\gtrsim2.5$
for a survey with $M_{lim}=20$ mag) \citep{Kochanek93a}. Thus, we shall 
disregard corrections of this nature, but remark that the necessity of
including this correction depends on the survey characteristics.

All these effects are somewhat connected, especially the magnitude bias,
angular separation and dynamic range effect \citep{Kochanek93a,Kochanek93b},
but since we are considering a hypothetical survey, we do not need to consider
a detailed selection function. Hence we will follow the simplified approach
of \citet{Cooray1}, and write the probability that a source at redshift
$z$ will be lensed and detected as
\be
p(z)=\tau B(<m)f(\Delta\theta)\;,
\ee
where $B(<m)$ is the average magnification bias for the survey, which will be
defined, and $f(\Delta \theta)$ is the angular selection function. Since
we found no information on the expected magnitude difference limit for the 
SDSS we choose just to consider the correction due to
the angular resolution. The ignored correction should be small ($\approx0.9997$
as estimated in \citet{Cooray1} for $\Delta m<2.5$, see also
\citet{Kochanek93a}), hence ignoring this effect will not yield a large
error.

Considering the assumptions commonly used in the literature (see, for instance,
\citet{FT}), the magnitude bias at a given magnitude level, 
$B(m)$, is given by
\be
B(m)=\frac{1}{N_Q(m)}\int_0^\infty N_Q(m+\Delta)P(\Delta)d\Delta\;,
\ee
where $N_Q(m)$ is the intrinsic quasar number counts, $\Delta=2.5\log
\mathcal{A}$, $\mathcal{A}$ being the total magnification produced by the lens,
and $P(\Delta)d\Delta$ is the probability distribution of $\Delta$. The
probability distribution of the magnification $\mathcal{A}$ is given by
\citep{TOG}
\be
P(\mathcal{A})d\mathcal{A}=\frac{8}{\mathcal{A}^3}d\mathcal{A}\qquad 
\mathcal{A}\ge2\;,
\ee
or alternatively
\be
P(\Delta)d\Delta=7.37\times10^{-0.8\Delta}d\Delta \qquad \Delta\ge0.75\;.
\ee

We shall use the quasar number counts from \citet{Hartwick},
\begin{equation}\label{qnc}
N_Q(m_b)=\left\{\begin{array}{ll}
(10/\square^\textrm{o})10^{0.86(m_B-19.15)}&m_B\le19.15\;,\\
(10/\square^\textrm{o})10^{0.28(m_B-19.15)}&m_B\ge19.15\;.
\end{array} \right.
\end{equation}
The $\square^\textrm{o}$ represents the number of sources per square degree of
sky. Note that the normalization of the number counts does not influence the
magnification bias. The magnification bias is given by \citep{FT}
\be\label{b(mb)}
B(m_B)=
\left\{
\begin{array}{lll}
59.5\times 10^{0.06(19.15-m_B)}-59.2&&m_B\le18.40\;,\\
2.5\times 10^{0.58(19.15-m_B)}&18.40<&m_B< 19.15\;,\\
2.50&&m_B\ge19.15\;.
\end{array}\right.
\ee

Since we are working with a hypothetical data set, we must average $B(m)$
over the observed magnitude distribution. The average bias for a survey whose 
magnitude limit is $m_L$, $B(<m_L)$, is then
\be
B(<m_B)=\frac{\int_{-\infty}^{m_L}N_Q(m_B)B(m_B)dm_B}{\int_{-\infty}^{m_L}
N_Q(m_B)dm_B}\;.
\ee

For the SDSS, \citet{CoorayHut} used a limiting magnitude of $m_L=21$ mag,
although the survey has a limiting magnitude of $m_L=23$ mag. We shall follow
this suggestion and also use $m_L=21$ mag, which corresponds to an
average magnification bias factor of $B(<m_L)=2.89$.

Moreover  we will use a step  angular selection function, such that the lens
is detected if the angular separation is between $0.1''$ and $6''$ , being 
undetected otherwise. For a SIS model, the fraction of lenses in this range
is $0.901$ (using the results from \citet{Kochanek93b,FT}). We shall
consider a simplified selection function such that all lenses within this
angular separation range will be recovered and write $f(\Delta \theta)=0.901$.
The angular limits used are very optimistic, but this provides an estimate of
how precise the survey must be to obtain sensible results.

Clearly, what is observed in strong gravitational lensing statistics is the
fraction of lensed sources relative to the overall number of sources. 
Consider that $N_{Total}$ sources (such as quasars or galaxies) are found.
To study the effect that the size of the survey will have on the precision
of the test, we consider three test models, with $N_{Total}=25000,10^5,
10^6$  sources (the SDSS survey is expected to yield the spectra of 
$10^5$ quasars, and about $10^6$ optical quasars. The 2dF survey expects to
find 25000 QSO), spread uniformly in 40 redshift bins of width $\Delta z=0.1$,
starting from $z_{min}=0$ until $z_{max}=4$.  In this model there will be 
$n=N_{Total}/40$ sources in each redshift bin. The effect of considering a
coarser angular resolution may be seen as having less sources at each redshift.
Thus, the $10^6$ case may be seen as the best scenario. A more realistic
angular selection function would give results that are between the $10^5$ and
the $10^6$ sources cases.

Consider now a fixed redshift bin $z_i$. For this redshift bin $n=N_{Total}/40$
sources are picked, each one with an independent probability of
being lensed given by $p(z_i)$. The number of lensed 
sources with redshift $z_i$, $N_{L}(z_i)$, is then a binomial random 
variable with parameters $n$ and $p(z_i)$. Since $n$ is large, and 
$p(z_i)$ is very small, one may use the Poisson PDF with average 
\be
N_{exp}(z_i)= np(z_i)\;.
\ee

\subsection{\label{section4C}Building Confidence Regions.}

In order to build up confidence regions we employ the Fisher information matrix
method (see e.g. \citet{Tegmark1997}), which provides an estimate of the
inverse covariance matrix when the likelihood function of the random variable
is known.

The Fisher matrix F is given by:
\be
F_{ij}=-\left<\frac{\partial^2\ln L}{\partial \theta_i\partial \theta_j}
\right>_{\bf x}
\ee
where $L$ is the likelihood of observing the data set ${\bf x}$ given the
parameters $\theta_1,\dots,\theta_\nu$, and the average is being computed 
at the point corresponding to the true set of parameters $\bm\Theta_{true}$.
The Fisher information matrix is the expectation value of the inverse of
the covariance matrix at the maximum likelihood point, and it gives us a
measure of how fast (on average) the likelihood function falls off around the
maximum likelihood point. Assuming the Poisson statistics as a good description
of the expected number of lenses as function of redshift, the  likelihood
function is written as
\beqa
L[n_L(\Delta z_1),n_L(\Delta z_2),...,n_L(\Delta z_{max})]=
\nonumber\\
=\prod_{z=0}^{z_{max}}\frac{e^{-N_{exp}(\Delta z_i)}
N_{exp}^{n_L(\Delta z_i)}}{n_L(\Delta z_i)!}\;.
\eeqa
Taking the logarithm and calculating the derivatives one obtains 
\citep{CoorayHut}
\be
F_{ij}=\sum_{\Delta z}\frac{1}{N_{exp}(z,\Delta z)}\frac{\partial N_{exp}(z,
\Delta z)}{\partial \theta_i}\frac{\partial N_{exp}(z,\Delta z)}{\partial
\theta_j}\;.
\ee

Confidence regions are built by imposing a confidence limit on the
likelihood function. To build these regions using the Fisher matrix one starts
by expanding the logarithm of the likelihood in a Taylor series around the
maximum likelihood  point. Since this corresponds to  a maximum of $L$, the
first derivative vanishes and one finds
\be
2\ln(L_{max}/L)=\delta{\bm \theta}F\delta{\bm \theta}\;,
\label{EllipseEquation}
\ee
where $\delta{\bm \theta}=(\theta_1-\theta_{1,max};...;\theta_\nu-\theta_{\nu,
max})$. By choosing a range of values of $2\ln(L_{max}/L)$ corresponding to the
desired confidence level, a second order equation in the $\delta \theta_i$ is
obtained which can be solved to obtain the desired confidence regions.
It can be shown that in the large $n$ limit, $2\ln(L_{max}/L)$ has a $\chi^2$
distribution with $\nu$ degrees of freedom \citep{Kochanek93b}. For our study
$\nu=2$, and the 68.3\%CL are found by imposing $2\ln(L_{max}/L)=2.3$, while
the 95\%CL regions are found by imposing  $2\ln(L_{max}/L)=5.99$.

It should be pointed out that this population model is not very realistic since
fewer sources are found at low redshift due to the smaller volume involved
and, on the other hand, fewer sources at high redshift as only very bright
sources are visible. Considering a more realistic distribution should not
change the results appreciably though, as one finds more sources at the
medium redshift range, thus with a higher lensing rate, but on the other,
fewer sources are found at high redshift, which results in an overall lower
lensing rate. These two competing effects should thus cancel each other, and 
produce a small net effect.

\subsection{\label{LensingResults}Results.}

\begin{figure}
\centering
\plotone{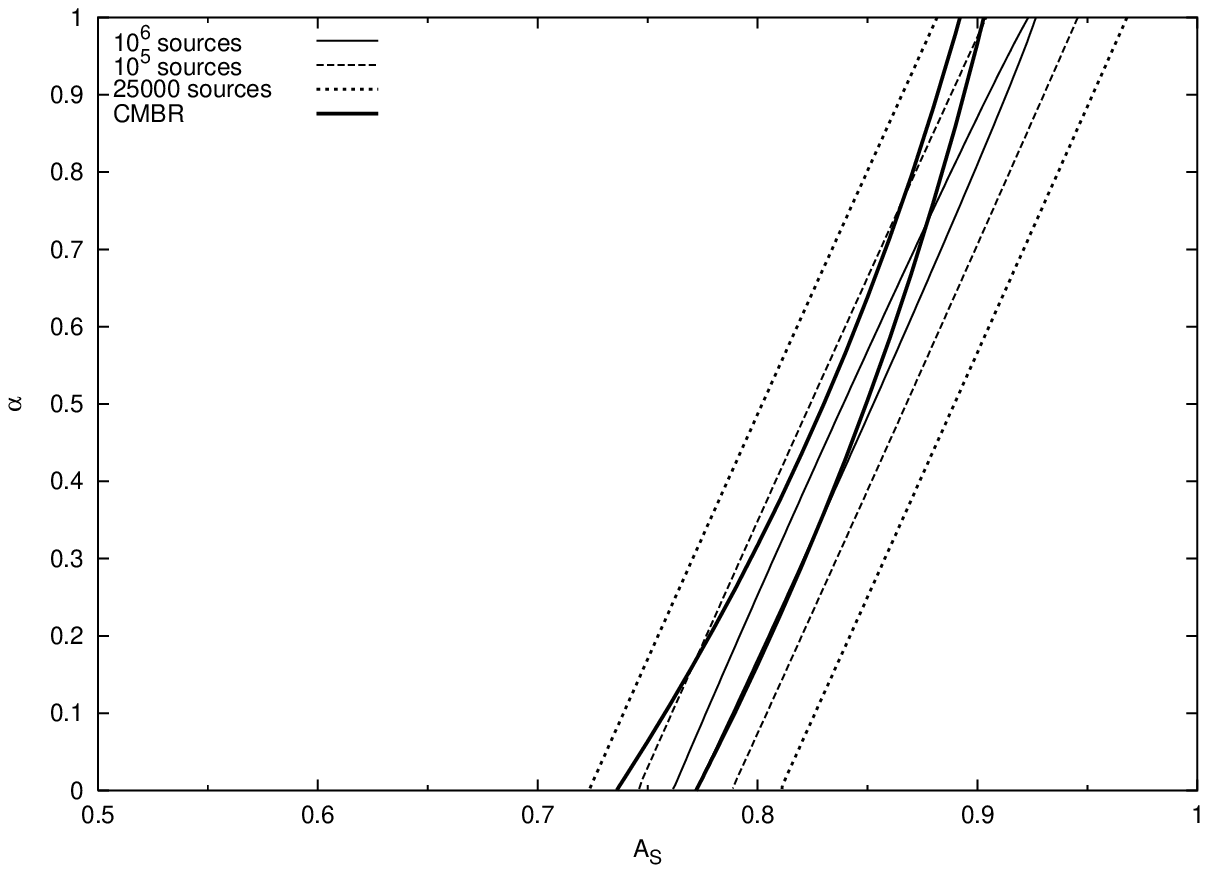}\\
\caption{Lensing 68\% confidence regions for Model I. The thick lines 
are the CMBR constraints from \citet{Bento2002b}.}
\label{Lensing68cl}
\end{figure}

\begin{figure}
\centering
\plotone{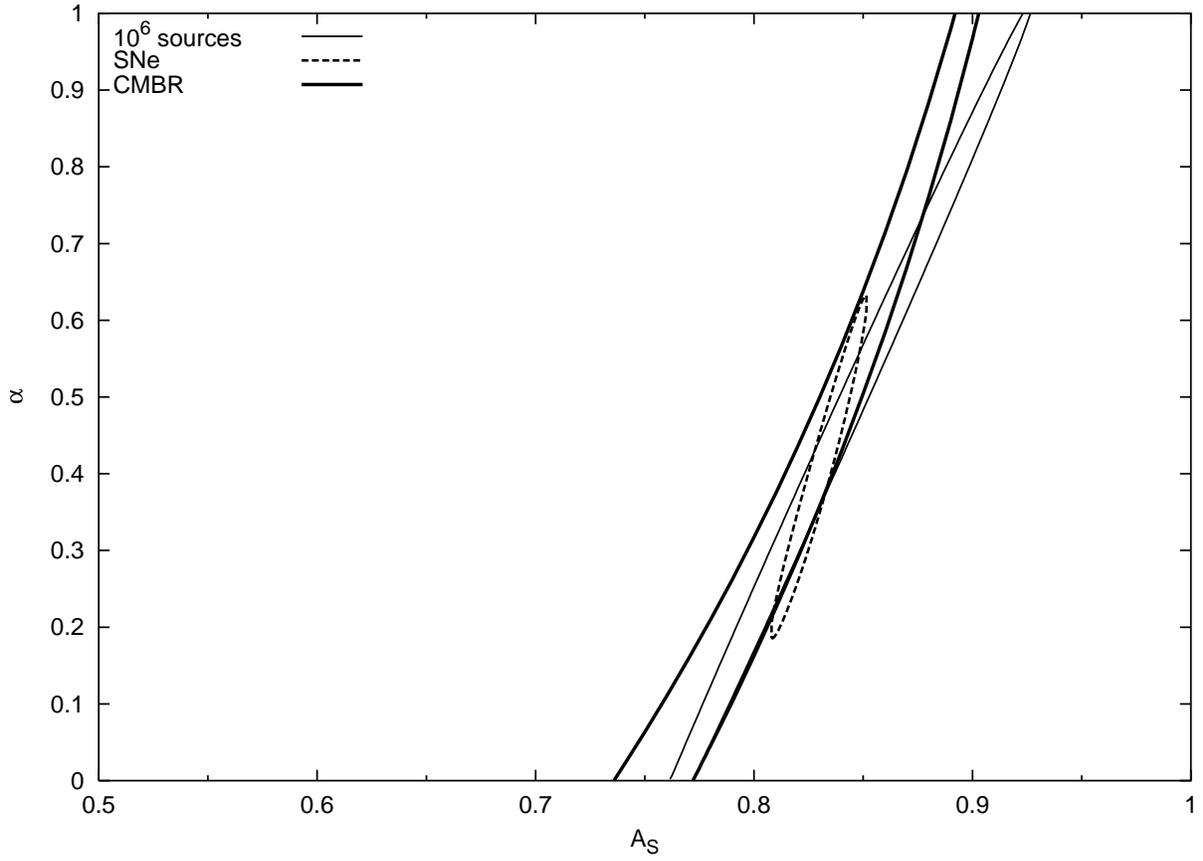}\\
\caption{Joint 68\% confidence regions for Model I from lensing, SNe and
CMBR constraints from \citet{Bento2002b}. The SNe confidence region is computed
assuming an exact $\mathcal{M}$}
\label{lplusSN68CL}
\end{figure}

\begin{figure}
\centering
\plotone{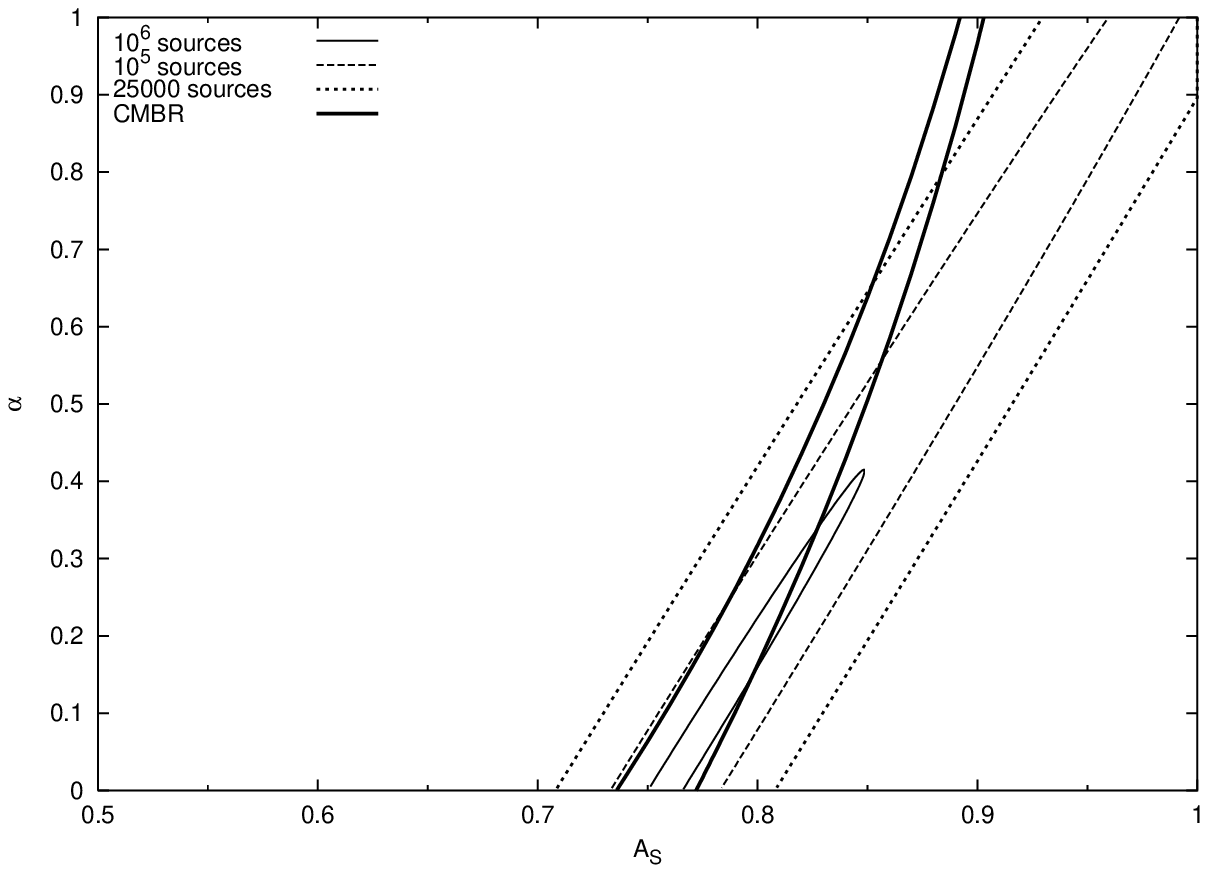}\\
\caption{Lensing 68\% confidence regions for Model II. The thick lines are
the CMBR constraints from \citet{Bento2002b}.}
\label{lcdm_Lensing68cl}
\end{figure}

\begin{figure}
\centering
\plotone{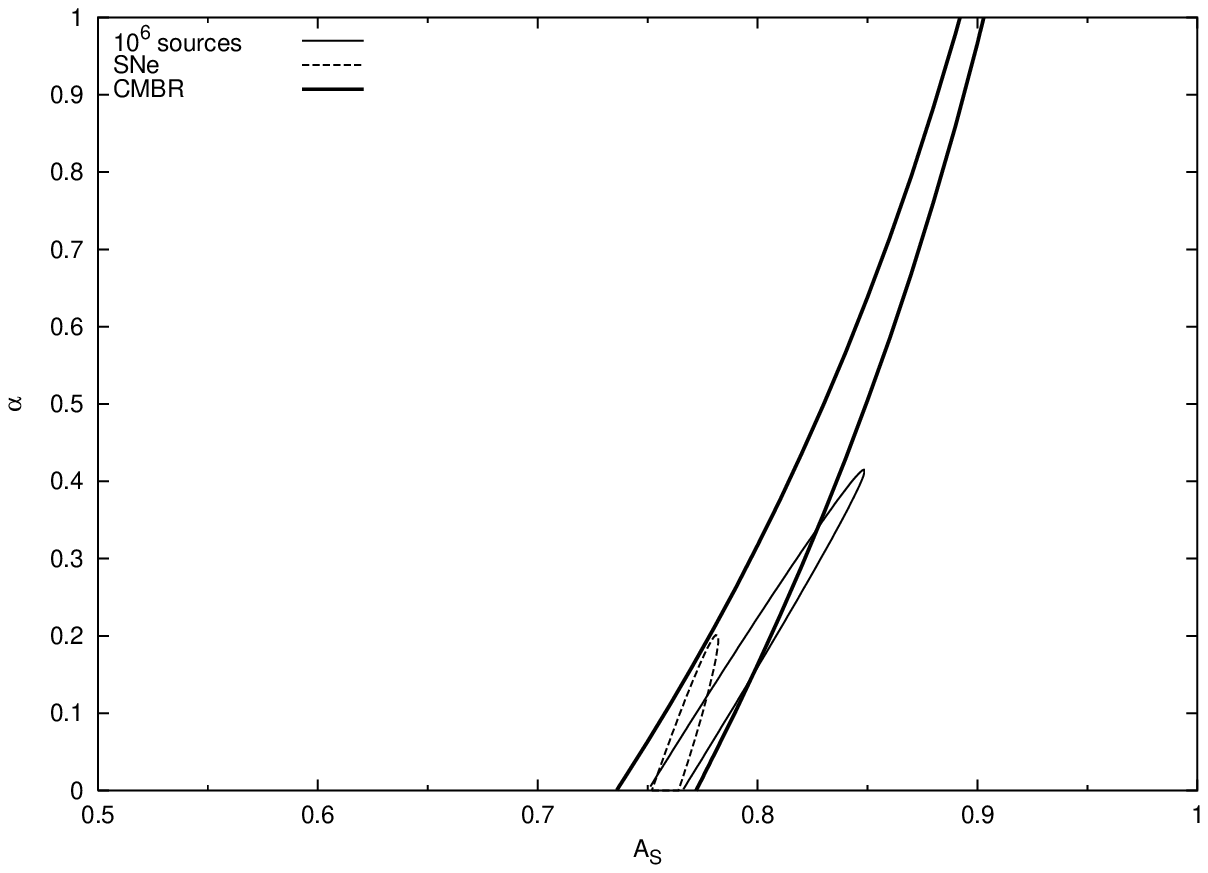}\\
\caption{Joint 68\%CL confidence regions for Model II using both 
SNe, gravitational lensing statistics and CMBR constraints from 
\citet{Bento2002b}. The SNe confidence region is computed assuming an exact
$\mathcal{M}$} 
\label{lcdmPlusSN68cl}
\end{figure}

\begin{figure}
\centering
\plotone{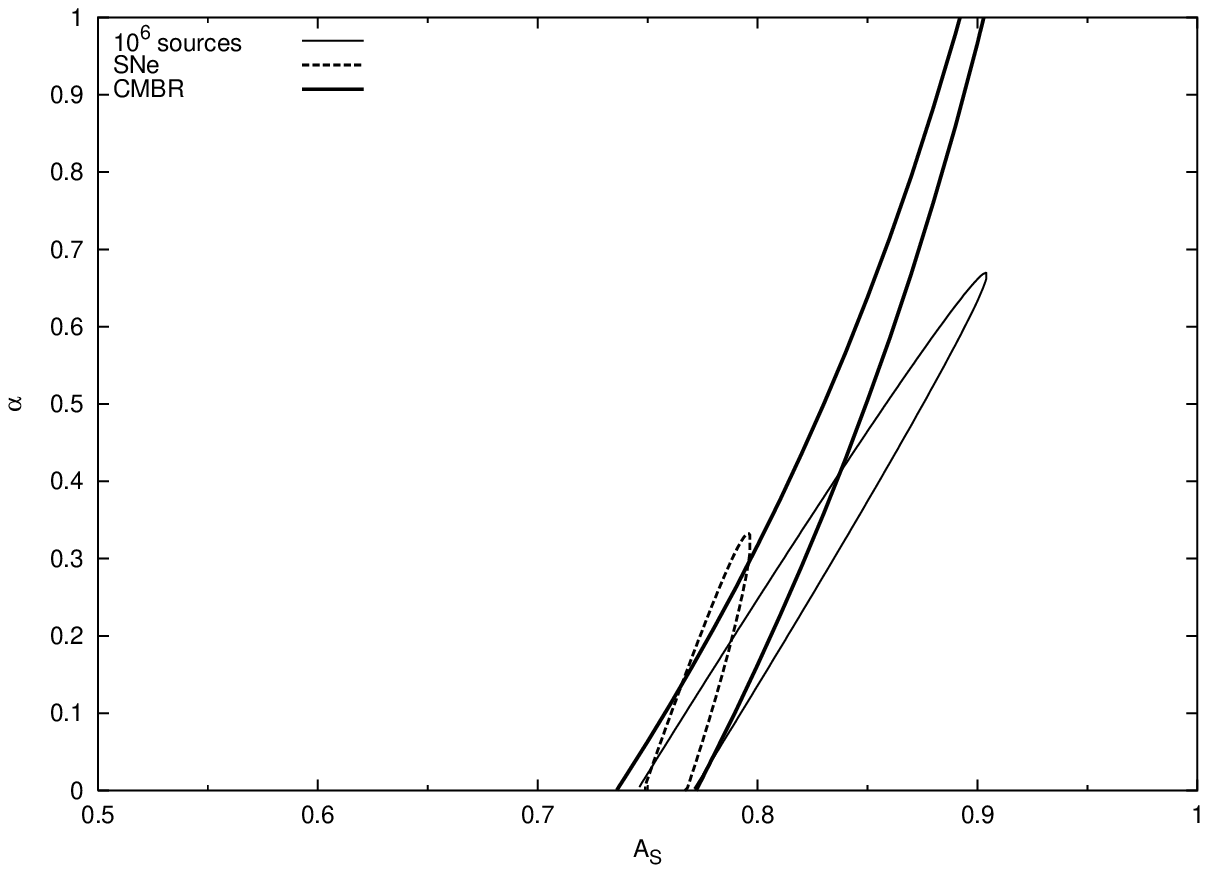}\
\caption{Joint 95\%CL confidence regions for Model II using both 
SNe, gravitational lensing statistics and  CMBR constraints from 
\citet{Bento2002b}. The SNe confidence region is computed assuming an exact
$\mathcal{M}$} 
\label{lcdmPlusSN95cl}
\end{figure}

The confidence regions that were built are shown in Figures
\ref{Lensing68cl}-\ref{lcdmPlusSN95cl}. Since their implications are quite
different, the results for each fiducial model are analysed separately. 
We consider strong lensing statistics as probe by themselves, and together
with CMBR peak location and SNAP constraints. This goes beyond the analysis
of \citet{Makler} where only the SNAP data was considered, and the analysis 
of \citet{Bento2002b,Bento2003}, where only the CMBR peak location was analised.

\subsubsection{Model I.}


The gravitational lensing statistics test by itself is unable to yield
important constraints on the parameter space of the GCG model, and, at most
allows asserting that $0.76\lesssim A_s\lesssim0.94$, with no limits on
$\alpha$. Even considering the CMBR peak position data from Archeops and
BOOMERANG only an upper value  $\alpha\lesssim0.7$, and
$0.76\lesssim A_s\lesssim0.87$ is found, as can be seen in Figure
\ref{Lensing68cl}.

Thus, combining CMBR constraints from \citet{Bento2002b}, lensing statistics
and SNAP data (Figure \ref{lplusSN68CL}) does not improve the limits found in
Section 3.4. There is only a marginal improvement in the upper range of the
allowed values of
$\alpha$. Therefore it is expected that, if the limits imposed
in \citet{Bento2002b} are accurate, SNe distance measurements and
strong gravitational lensing statistics we will not be able to substantially
improve the present constraints, at least in the near future.

\subsubsection{Model II.}

For the second fiducial model, the results are more promising. As may be seen
in Figure \ref{lcdm_Lensing68cl}, the search
for lenses in the SDSS will allow to impose a fairly tight parameter range
$\alpha\lesssim0.3$, and $0.75\lesssim A_s\lesssim0.85$. For smaller lens 
surveys, these results together with CMBR peak location data may yield some
constraints. It is particularly interesting to note that a survey with the size
of the 2dF survey might be able to place some interesting upper limits on
$A_s$ and $\alpha$ in the very near future, although it should be realized that
using a more careful treatment of the selection effects might broaden the
confidence regions.

The most interesting results appear when we use both SNAP data and
lensing statistics for a large survey such as the SDSS (Figures
 \ref{lcdmPlusSN68cl} and \ref{lcdmPlusSN95cl}). We are close to 
determining the model, as they will allow to constrain the parameters to be in
the range $\alpha\lesssim0.1$ and $0.75\lesssim A_s\lesssim0.77$ at 68\% CL, or
 $\alpha\lesssim0.2$ and $0.75\lesssim A_s\lesssim0.79$ at 95\% CL. This 
means that SNAP data together with strong lensing statistics from a large
precise survey should be able to rule out the CGC model as an unification
scheme between dark matter and dark energy.


\subsection{Conclusions and Outlook.}

In this paper we have studied how the SNAP data and the data from future quasar
surveys might be used to constrain a generalized Chaplygin equation of state.
Our conclusions depend on the models under consideration.

If the fiducial model is the model favored by available SNe \citep{Makler}, 
CMBR peak location \citep{Bento2002b} and age estimates of distance sources
\citep{Dev2} (Model I in Table \ref{FiducialModels}), then the SNAP data will
be able to definitely rule out both the pure Chaplygin case, as well as  the
$\Lambda CDM$ model, but it will not be able to provide tighter constraints
than those already available. This is particularly so without a precise
determination of the Hubble constant, $H_0$, and of the SNe absolute
magnitudes, $M$ in Eq. (\ref{mz}). In fact, to completely rule out the
$\Lambda CDM$ model, a good determination of these quantities is 
absolutely required.

Using the available data on the CMBR peaks positions from BOOMERANG 
\citep{Boomerang} and from Archeops \citep{Archeops}, a slight improvement 
in the allowed parameter range is expected, but most likely to verify the
 range of
parameters determined in \citet{Bento2002b}, namely $0.81\lesssim A_s
\lesssim0.85$ and $0.2\lesssim\alpha\lesssim0.65$ at 68\% CL, or $0.15\lesssim
\alpha\lesssim0.75$ and $0.79\lesssim A_s \lesssim0.87$ at 95\% CL. Of course
that the situation will improve greatly with data from WMAP, especially with a
more precise determination of the location of the acoustic peaks
\citep{Bento2002b}.

Lensing statistics by themselves will not be very helpful in constraining the 
allowed range of parameters, but together with SNAP data and CMBR constraints,
the allowed parameter range is slightly tighter, yielding $0.81\lesssim A_s
\lesssim0.85$, and $0.2\lesssim\alpha\lesssim0.6$ at 68\% CL. We therefore
conclude that in order to achieve more precise limits on our fiducial Model I,
the best strategy is to aim for better determinations on the location of the
CMBR acoustic peaks. If the Universe is best described by Model I, the data
from SNAP or from the search for lenses in large quasar surveys will  only be
able to verify the present day limits, and only marginally to improve them.

Turning now to the $\Lambda CDM$ fiducial model (Model II in Table
\ref{FiducialModels}), the results are much better. Again, an accurate
determination of the zero point  $\mathcal{M}$ is required in order to achieve
some precision. With such a determination one might rule out 
models with $\alpha\gtrsim0.2$ at 68\% CL, or with $\alpha\gtrsim0.35$ at 
95\% CL.  Using the CMBR peaks locations, marginal improvements may be
obtained. The best results arise from the use of SNe and lensing statistics,
where one comes close to determining the model, obtaining the allowed range,
$\alpha\lesssim0.1$ and $0.75\lesssim A_s\lesssim0.77$ at 68\% CL, and $\alpha
\lesssim0.2$ and $0.75\lesssim A_s\lesssim0.79$ at 95\% CL. The lensing results
should be regarded with care, since our best case scenario is overtly
optimistic.

Therefore, if one considers a two component Universe made up of baryons and the
GCG, future data from SNAP and from large QSO surveys, such
as the SDSS, will be able to test the GCG model, and rule out a
possible unification scheme between dark energy and dark matter it provides.
The question of how well we will be able to constrain the parameter space
depends on the model that is ruled out.

\subsubsection{Assumptions made for strong lensing statistics.}

As a final note, we mention that caution should be exerted when considering
results from gravitational lensing statistics, as there are several systematic
effects that we have not explored. The inclusion of a core radius on the galaxy
model, for instance, will greatly suppress the lensing optical depth
\citep{Hinshaw}, but on the other
hand it will enhance the magnification bias \citep{Kochanek96}. This partly
compensates the lower lensing probability and produces a small net effect.
Also, the lack of central images in observed gravitational lenses indicates
that the core radii will be smaller than $200$ pc \citep{Wallington}. These
results also agree with \citet{Kochanek96}, and \citet{FT} where strong lensing
statistics were used to constrain several systematic uncertainties. Therefore,
the error from ignoring the core radius appears to be negligible. However,
since one is dealing with an unprecedented degree of precision, uncertainties
will inevitably affect the results.

Effects due to galactic evolution have been studied in \citet{Mao},
\citet{MaoK},and \citet{Rix}, with  the conclusion that evolution will only
affect the statistics substantially if it is important  at $z\lesssim1$. For a
vacuum dominated
Universe this should not be a problem since galaxies form earlier that for a
matter dominated Universe \citep{Carrol}. Still, evolutionary studies of
galaxies have not reached conclusive results. For instance, results from the
Canada-France Survey \citep{Lilly} indicate that the comoving number density of
sources is independent of redshift. On the other hand, results from the Hubble
Deep Field depend on the methods used to estimate the redshift distribution of
sources. In \citet{Mobasher} it is argued that there are no signs of evolution,
while \citet{Sawicki} indicate that there is some evolution, although
not a dramatic one, from  $0.2<z<0.5$ until $1<z<2$. On the other hand, a more 
dramatic evolution from $z=2.2$ until $z=0.6$ is argued in \citet{Gwin}.

In calculating the parameter $F$ in Eq. (\ref{F}), we must relate the
velocity dispersion of matter in the galaxy to its luminosity. 
The problem of estimating the dark matter velocity 
dispersion $\sigma_{DM}$ from the luminous matter in elliptical galaxies
arises, and is of great importance in gravitational lensing statistics, since a
22\% change of $\sigma_\|$ in Eq. (\ref{F}) leads to a 125\% change in the 
expected number of  lenses and a 50\% change in image separations. Here we
assumed that  the velocity dispersion of dark matter is the same as the
luminous matter, as considered in  \citet{Kochanek91,Kochanek93a,Kochanek93b,
Kochanek96}, \citet{Chiba}, \cite{Cooray1}, \cite{CoorayHut}, and 
\citet{CoorayAl}. This equality was advocated in \citet{Kochanek93b} and
\citet{Kochanek96}.
Other studies have suggested that a correction of order $(3/2) ^2$ should be
included in the dark matter velocity dispersion \citep{TOG}. This correction
is most likely inaccurate since  it was computed using the global virial
theorem, which is not valid for power law profiles that diverge at the origin
or at infinity. Fitting the observed
velocity dispersion as function of radius to tracer models of the isothermal
sphere profile, it was concluded that $\sigma_{DM} \sim v_{los}$ for regions
close to the center of the galaxies \citep{Kochanek93b}. This was
verified  by lensing statistics also, in \citet{Kochanek94}. This 
conclusion is also supported by detailed dynamical modeling of early type
galaxies \citep{Breimer1993, Franx1993,Kochanek94}. 
There is also a 30\% uncertainty 
in the value of $F$, which we did not take into account. This uncertainty is 
likely to become smaller once more data on galaxies become available.

In our estimates, effects such as reddening and the luminosity of the lensing
galaxy \citep{Kochanek91} were ignored as there is still not enough
quantitative knowledge of their results, but it is known that for a faint
survey, such as the SDSS, effects due to the luminosity of the lensing galaxy
must be taken into account.

\acknowledgments
It is a pleasure to thank A.A. Sen, and L. Teodoro for the discussions on 
gravitational lensing. O.B. acknowledges the partial support of
Funda\c c\~ao para a Ci\^encia e a Tecnologia (Portugal)
under the grant POCTI/1999/FIS/36285.


\clearpage

\begin{table}
\caption{SNAP specifications for a two year period of observations.}
\begin{center}
\label{SNAPerrors}
\begin{tabular}{ccccc}
\tableline
\tableline
Redshift Interval &z=0-0.2&z=0.2-1.2&z=1.2-1.4&z=1.4-1.7\\
\tableline
Number of SNe &50&1800&50&15\\
\tableline
\end{tabular}
\end{center}
\end{table}

\clearpage
\begin{deluxetable}{lcccc}
\tablewidth{0pt}
\tablecaption{Parameters for the considered fiducial models.
\label{FiducialModels}}
\tablehead{
\colhead{} &  \colhead{$\Omega_b$} & \colhead{$\Omega_{ch}$} &
\colhead{$A_s$} & \colhead{$\alpha$}}
\startdata
Model I & 0.05 & 0.95 & 0.83 & 0.4 \\
Model II & 0.05 & 0.95 & 0.758 & 0.0\\
\enddata
\end{deluxetable}

\clearpage



\end{document}